\documentclass{emulateapj}

\slugcomment{Final (accepted) version from March 12, 2012}
\shorttitle{HVCs and their connection to QSO absorbers}
\shortauthors{Philipp Richter}

\begin{document}

\title{
Cold gas accretion by high-velocity clouds\\
and their connection to QSO absorption-line systems
}

\author{Philipp Richter
}
\affil{Institut f\"ur Physik und Astronomie, Universit\"at Potsdam,\\
Haus 28, Karl-Liebknecht-Str.\,24/25, 14476 Golm (Potsdam),
Germany} 
\affil{Leibniz-Institut f\"ur Astrophysik Potsdam (AIP),
An der Sternwarte 16, 14482 Potsdam, Germany}
\email{prichter@astro.physik.uni-potsdam.de}

%

\begin{abstract}

We combine H\,{\sc i} 21cm observations of the Milky Way, M31, and the
local galaxy population with QSO absorption-line measurements to geometrically 
model the three-dimensional distribution of infalling neutral gas clouds 
(``high-velocity clouds'', HVCs) in the extended halos of low-redshift galaxies. 
We demonstrate that the observed distribution of HVCs around the Milky Way and M31
can be modeled by a radial exponential decline of the mean H\,{\sc i} 
volume filling factor in their halos. Our model suggests a characteristic 
radial extent of HVCs of $R_{\rm halo}\sim 50$ kpc, a total H\,{\sc i}
mass in HVCs of $\sim 10^8$\,$M_{\sun}$, and a neutral-gas-accretion
rate of $\sim 0.7\,M_{\sun}$\,yr$^{-1}$ for M31/Milky-Way-type galaxies.
Using a Holmberg-like luminosity scaling of the halo size of galaxies we 
estimate $R_{\rm halo}\sim 110$ kpc for the most massive galaxies.
The total absorption-cross section of HVCs at
$z\approx 0$ most likely is dominated by galaxies with total H\,{\sc i}
masses between $10^{8.5}$ and $10^{10}$\,$M_{\sun}$.
Our model indicates that the H\,{\sc i} disks of galaxies and their
surrounding HVC population can account for $30-100$ percent of 
intervening QSO absorption-line systems with log $N$(H\,{\sc i}$)\geq
17.5$ at $z\approx 0$. We estimate 
that the neutral-gas accretion rate density of galaxies at low redshift
from infalling HVCs is $dM_{\rm HI}/dt/dV \approx 0.022\,M_{\sun}$\,yr$^{-1}$
\,Mpc$^{-3}$, which is close to the measured star-formation rate density in the 
local Universe.  HVCs thus may play an important role for the on-going formation
and evolution of galaxies.

\end{abstract}

%
 
%

\section{Introduction}

One crucial aspect of galaxy formation and evolution concerns
the continuous infall of intergalactic gas onto galaxies.
While it is clear that galaxies do accrete substantial amounts
of gas from intergalactic space to power star formation, the
exact way of {\it how} galaxies get their gas is still
a matter of debate. In the conventional sketch of galaxy
formation and evolution gas is falling into a dark matter (DM)
halo and then is shock-heated to approximately the
halo virial temperature (a few $10^6$ K, typically), 
residing in quasi-hydrostatic equilibrium with the 
DM potential well (Rees \& Ostriker 1977). The gas 
then cools slowly through radiation,
condenses and settles into the center of the potential 
where it forms stars as part of a galaxy (`hot mode' of
gas accretion). It has been argued, however, 
that for smaller DM potential wells the infalling gas may radiate
its acquired potential energy at much lower temperatures
($<10^{5.5}$ K, typically), so that one speaks of the `cold mode' 
of gas accretion (e.g., White \& Rees 1978). For the 
cold mode of gas accretion the star-formation rate of 
the central galaxy is directly coupled to its gas-accretion
rate (White \& Frenk 1991). 
Numerical simulations indicate that for individual 
galaxies the dominating gas-accretion mode depends 
on the mass and the redshift
(e.g., Birnboim \& Dekel 2003; Kere$\check{s}$ et al.\,2005). 
The general trend for $z\approx0$
is that the hot mode of gas accretion dominates for massive
galaxies with DM-halo masses $>10^{12}\,M_{\sun}$,
while the cold accretion mode dominates for galaxies
with smaller DM-halo masses (e.g., van\,de\,Voort et al.\,2011). 

Independently of the theoretically expected gas-accretion 
mode of galaxies it is known since a long time 
that galaxies at low and high $z$ are surrounded by large 
amounts of neutral and ionized gas that partly originates in the 
IGM. This material is complemented by neutral and ionized 
gas that is expelled from the galaxies as part of galactic 
fountains, galactic winds, and from merger processes (see, e.g., 
Richter 2006 for a review). Because the interplay between these
circumgalactic gas components is manifold and the gas
physics of such a turbulent multi-phase medium is 
complex, the circulation of neutral and ionized
gas in the inner and outer halos currently
cannot be modeled in full detail in hydrodynamical simulations.
To improve current models of galaxy-evolution models it is
of imminent importance to quantify the amount of cool, neutral
gas in and around galaxies from observations and search
for observational strategies to separate metal-deficient
infalling intergalactic gas from metal-enriched gaseous material 
that is circulating in the circumgalactic environment of 
galaxies as a result of fountain processes and galaxy mergers.

In the Milky Way and other, very nearby spiral galaxies 
(e.g., M31), the infall of neutral gas onto to the disks
can be observed directly by H\,{\sc i} 21cm observations of 
extraplanar gas clouds that move through the halos of these
galaxies. For the Milky Way, the so-called ``high-velocity clouds''
(HVCs) represent the prime candidates for neutral gas 
that is being accreted onto the Milky Way disk. 
HVCs represent high-latitude gaseous structures (located 
in the Galactic halo) observed in H\,{\sc i} 21cm emission
at high radial velocities, $|v_{\rm LSR}| > 100$ km\,s$^{-1}$
(e.g.; Wakker \& van Woerden 1998; Richter 2006). 
Halo clouds with somewhat smaller radial velocities in 
the range $|v_{\rm LSR}|=50-100$ km\,s$^{-1}$ are commonly
referred to as ``intermediate-velocity clouds'' (IVCs). 
Throughout this paper, we will use the expression ``HVC''
for all neutral halo clouds (including IVCs), if not
otherwise stated. The total neutral gas mass of the 
Milky Way's HVC population is on the order
of $10^8\,M_{\sun}$ and the total accretion rate of
neutral gas in the form of HVCs has been estimated to
be $\sim 0.5\,M_{\sun}$\,yr$^{-1}$ (e.g., Wakker et al.\,2007, 
2008; Wakker 2004).
Also M31 exhibits a population of neutral halo 
clouds in a similar mass range (for simplicity, hereafter also 
referred to as ``HVCs''; Thilker et al.\,2004). These 
observations, together with H\,{\sc i} 21cm measurements of 
other nearby galaxies (e.g., Sancisi et al.\,2007), imply
that HVCs represent a common phenomenon in the local Universe.

Another important method to study the gaseous outskirts
of galaxies and their relation to the cosmic web 
is the analysis of intervening absorption-line
systems in optical und ultraviolet (UV) 
spectra of QSOs (AGN). QSO absorption 
spectroscopy allows us to detect both neutral and ionized gas
in the intergalactic medium and the halos of galaxies 
over eight orders of magnitude in column density and
over more than 90 percent of the age of the Universe.
It is therefore a particularly sensitive method to
explore the multi-phase nature of circumgalactic 
gas and its origin (e.g., Bergeron \& Boiss\'e 1992; 
Steidel 1995; Charlton \& Churchill 1998; Ding et al.\,2005; 
Richter et al.\,2011).

In this paper, we combine H\,{\sc i} 21cm observations of the
HVC population of Milky Way and M31
with 21cm data of the local galaxy population and
QSO absorption-line measurements to study the three-dimensional
distribution of (partly) neutral gas structures in the extended halos of 
low-redshift galaxies. 
While several previous studies have linked
the Galactic HVC population to intervening QSO absorbers based
on various arguments (e.g., Blitz et al.\,1999;
Charlton, Churchill \& Rigby 2000; Pisano et al.\,2004;
Mshar et al.\,2007; Schaye, Carswell \& Kim 2007; Narayanan et al.\,2008;
Richter et al.\,2009; Stocke, Keeney \& Danforth 2010;
Ribaudo et al.\,2011), a detailed geometrical HVC model 
that connects key observables (column densities, covering fractions, 
absorber number densities) of HVCs and intervening absorbers
with the gas-accretion rates of galaxies at low redshift has not
been presented so far. The main goal of the present study is to
model the {\it radial} distribution of HVC analogs in halos as a 
function of galaxy mass and size, determine their absorption-cross 
section at $z\approx0$, and estimate the neutral-gas accretion rate 
of galaxies at low redshift.

We refrain from including H\,{\sc i} 21cm observations of
HVCs from other galaxies beyond the Local Group in our study, because 
these observations are strongly limited in sensitivity for detecting
diffuse neutral halo gas (e.g., Oosterloo et al.\,2007). In addition,
beam-smearing effects are known to smoothen out the true 
spatial distribution of individual halo clouds in more
distant galaxies, so that the observed filling factor of 
H\,{\sc i} 21cm in these systems does not provide a 
realistic estimate for the absorption-cross section of this gas.

\begin{deluxetable}{lrc}
\tablewidth{0pt}
\tablecaption{Properties$^{\rm a}$ of prominent Milky Way HVCs}
\tablehead{
\colhead{HVC name}  & \colhead{$f_{\rm HVC}^{\rm b}$} &
\colhead{$D^{\rm c}$}\\
\colhead{}          & \colhead{}              & \colhead{[kpc]}
}
\startdata
Complex A         & 0.01 & $8-10$    \\
Complex C         & 0.04 & $\sim 10$    \\
Complex H         & 0.01 & $>5$      \\
Magellanic Stream & 0.04 & $\sim 50$ \\
Complexes WA, WB  & 0.01 & $8-20$ \\
\enddata
\tablenotetext{a}{{\it References:} Wakker (2001,2004);
Wakker et al.\,(1999, 2007); B.P. Wakker (priv.\,comm.);
Thom et al.\,(2006, 2008); Gardiner \& Noguchi (1996)}
\tablenotetext{b}{Sky-covering fraction of HVC gas}
\tablenotetext{c}{Distance}
\end{deluxetable}

The paper is organized as follows: in $\S2$ we discuss
H\,{\sc i} 21cm observations of the HVC population of the 
Milky Way and M31. In $\S3$ we develop a simple model for the 
radial distribution of neutral gas in the halos of these
galaxies. In $\S4$ we generalize our HVC model for the 
local galaxy population, based on information on the 
H\,{\sc i} mass function of low-redshift galaxies and
the absorption-cross section of Damped 
Lyman $\alpha$ absorbers (DLAs) and Lyman limit systems 
(LLS) at $z\approx0$. The relation between HVCs and 
intervening QSO absorbers is discussed in $\S5$.
In $\S6$ we present the conclusions from our study.

%

\begin{figure*}[t!]
\epsscale{1.0}
\plotone{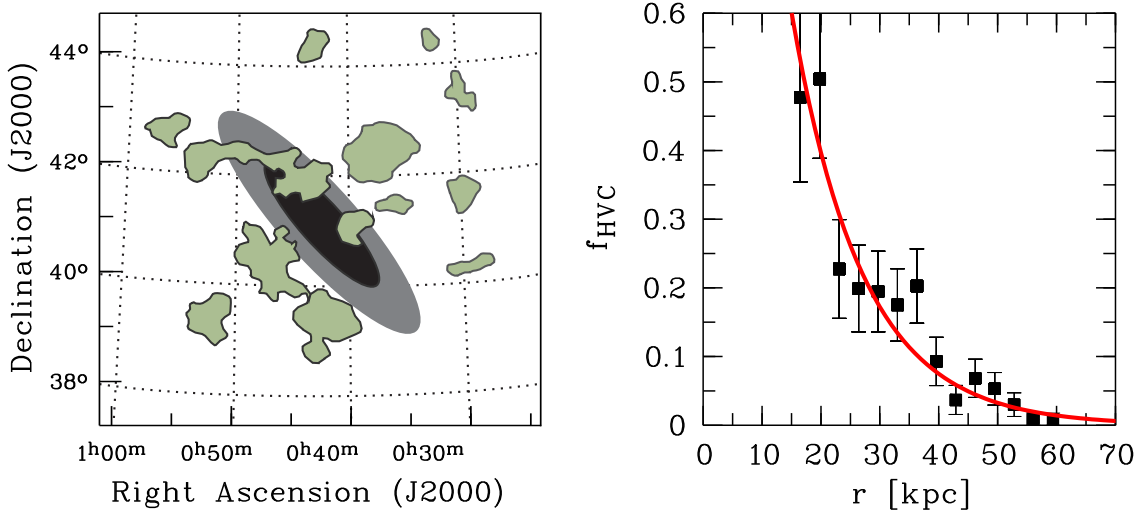}
\caption{
{\it Left:} distribution of HVCs (green) around M31, adopted from
21cm data of Thilker et al.\,(2004). The optical disk of M31 is indicated
with the black ellipse, the H\,{\sc i} disk is shown as gray-shaded
ellipse. Note that the shown HVC distribution is incomplete for small
radii ($r<15$ kpc) due to confusion of H\,{\sc i} halo gas with the
H\,{\sc i} disk of M31. {\it Right:} projected covering fraction of HVC gas
around the M31 disk, $f_{\rm HVC}$,
as a function of the projected radius, $r$, based on the HVC distribution
shown in the left panel. The covering fraction can be
fitted by an exponential in the form 2.1\,exp$(-r/12)$, as indicated
with the red solid line.
}
\end{figure*}

\section{High-velocity clouds in the halos of the Milky Way and M31}

\subsection{HVCs in the Milky Way}

The HVC population of the Milky Way has a total sky covering fraction 
of $f_{\rm c}\approx 0.30$ for column densities 
$N$(H\,{\sc i}$)\geq 7\times10^{17}$ cm$^{-2}$ and 
$f_{\rm c}\approx 0.15$ for column densities
$N$(H\,{\sc i}$)\geq 2\times10^{18}$ cm$^{-2}$ (Wakker 2004 and 
references therein). The largest Milky HVC is Complex C, which covers 
$\sim 1500$ square-degree on the sky ($f_{\rm c}\approx 0.04$). With 
a low metallicity of $\sim 0.15$ solar (e.g., Fox et al.\,2004; Richter
et al.\,2001) and a distance of $D\sim 10$ kpc (Wakker et al.\,2007; 
Thom et al.\,2008), Complex C most likely is a cloud 
that is being accreted from the IGM or from a 
satellite galaxy. The Magellanic Stream (MS) also covers an 
area of $\sim 1500$ square-degree, but most likely has a distance 
as large as $D\sim50$ kpc (Gardiner \& Noguchi 1996). The MS represents
a tidal feature expelled from the Magellanic Clouds as they move through
the extended Milky Way halo. With its large distance and its stream-like 
shape the MS is clearly distinct from most of the other Galactic HVCs, 
which predominantly are less extended and 
located at distances $<15$ kpc from the disk
(Wakker et al.\,1999, 2007, 2008; Thom et al.\,2006, 2008). 

Other prominent Galactic HVCs are Complex A, Complex H, the Anti-Center Cloud,
and Complexes WA$-$WE. Covering fractions and distances for some of these 
complexes (as far as known) are summarized in Table 1.
Because {\it accurate} distance information on the Milky Way HVCs is still
limited, the total H\,{\sc i} mass in HVCs and the H\,{\sc i} mass-accretion rate
ist not well constrained. The currently available information implies
$M_{\rm HI,HVC}\sim 3\times 10^8\,M_{\sun}$ and $dM_{\rm HI,HVC}/dt \sim
0.5\,M_{\sun}\,$yr$^{-1}$ (Wakker et al.\,2007; 2008; Wakker 2004). 
The H\,{\sc i} column densities in
HVCs follow a column-density distribution function (CDDF) in the form
$f(N_{\rm HI})\propto N_{\rm HI}^{-\beta}$ with $\beta=1.42$ 
for log $N$(H\,{\sc i}$)\geq 18$
(Lockman et al.\,2002).

\newpage

\subsection{The M31 HVC population}

As has been shown by Braun et al.\,(2009), M31 has an H\,{\sc i} disk
that extends to radii of $r\approx 30$ kpc for column densities above
log $N$(H\,{\sc i}$)\geq 20.3$. Beyond $r\approx 30$ kpc the H\,{\sc i} disk 
appears to be truncated (Braun et al.\,2009; Braun \& Thilker 2004).
The HVC population of M31 was studied 
in detail with the Green Bank Telescope (GBT) 
by Thilker et al.\,(2004). These authors detected more than 20 individual
HVCs around M31 with H\,{\sc i} column densities 
$\geq 5\times10^{17}$ cm$^{-2}$ and estimated a total H\,{\sc i} mass 
of the HVCs of $3-4\times 10^7\,M_{\sun}$ (see also Thilker, Braun \& 
Westmeier 2005). 

In the left panel of Fig.\,1 we show
the distribution of HVCs around the M31 disk, based on the 
GBT H\,{\sc i} 21cm contour map presented by Thilker et al.\,(2004).
The HVC population of M31 reaches out until $\sim 50$ kpc with
a strongly decreasing HVC covering fraction towards larger radii.
We assume $D=785\pm25$ kpc as distance for M31 
(McConnachie et al.\,2005), so that absolute distance estimates derived 
from angular coordinates are uncertain by $\sim 3$ percent.
Some interesting conclusions about the radial distribution
of neutral gas around M31 can be drawn from this HVC
distribution map. Fig.\,1, right panel, shows the radius-dependent 
projected covering fraction, $f_{\rm HVC}(r)$, of the HVC
population of M31 plotted against $r$ (indicated
by the filled boxes), where $r$ is the projected radius. 
To calculate $f_{\rm HVC}(r)$ we have resampled the M31 HVC
map of Thilker et al.\,(2004) and have transformed the ($\alpha,\delta$) 
coordinate system into a polar coordinate system with coordinates $r$ and 
$\theta$ centered on M31.
For each ring with radius $r$ and thickness $r\pm\Delta r$ 
the parameter $f_{\rm HVC}(r)$ then was derived by comparing 
the area covered by HVC gas with the total ring area.
The error bars for $f_{\rm HVC}(r)$ shown in Fig.\,1 have 
been calculated assuming Poisson-like statistics.
Starting from $f_{\rm HVC}\approx 0.5$
at $r=15$ kpc the covering fraction decreases to values less than 
$0.05$ for radii larger than $r=45$ kpc. 
This trend for $f_{\rm HVC}(r)$ can be fitted by an exponential 
in the form $f_{\rm HVC}(r)=x$\,exp$\,(-r/y)$ with $x=2.1\pm0.2$ and
$y=12.0^{+0.7}_{-0.5}$, as shown by the solid red line in the right
panel of Fig.\,1. 
An exponential fit to the extraplanar H\,{\sc i} features of M31
was also favored by Braun \& Thilker (2004), who analyzed
lower-resolution 21cm data of M31 from the {\it Westerbork Synthesis
Radio Telescope} (WSRT).

For the inner regions of the M31 halo at $r<15$ kpc no HVC data
are available. As discussed by Thilker et al.\,(2004), 
this does not imply that this region is devoid of HVC material.
The lack of data for the disk-halo interface region at 
$r<15$ kpc (Fig.\,1, left panel) rather indicates the 
incompleteness of the HVC map for small radii because of 
the confusion of neutral halo gas with the H\,{\sc i} disk of 
M31 together with the stringent selection criteria defined by
Thilker et al.\, to unambiguously identify HVC features. 
In the Milky Way, the disk-halo interface at $r<15$ kpc 
is filled with large amounts of neutral gas that gives rise
to 21cm emission at intermediate and high velocities (e.g.,
Wakker 2004). Nearby edge-on galaxies such as NGC\,891 
also exhibit large amounts of neutral gas in the disk-halo 
interface region extending several kpc above and below the disk
(see Sancisi et al.\,2008). 21cm measurements of NGC\,891 
show that the (projected) covering fraction of neutral gas 
is $f_{\rm HI}=1$ for vertical distances $d<10$ kpc to the 
midplane of the NGC\,891 disk (Oosterloo et al.\,2007).
If we extrapolate $f_{\rm HVC}$ for M31 to small radii using
the exponential defined above, $f_{\rm HVC}$ becomes unity
for $r\leq 9$ kpc, in line with the extraplanar gas distribution 
observed in NGC\,891. If we define $r_3$ as the radius 
beyond which the projected covering fraction of HVCs falls 
below 3 percent, we derive for M31 a value of 
$r_3=50\pm 6$ kpc. The $1\sigma$ error reflects the
uncertainties in the exponential parameters for
$f_{\rm HVC}(r)$.

\begin{deluxetable*}{lll}
\tabletypesize{\small}
\tablewidth{0pt}
\tablecaption{Model parameters for Milky Way/M31 HVCs}
\tablehead{
\colhead{Parameter}  & \colhead{Description} & \colhead{Value}\\
}
\startdata
\multicolumn{3}{c}{Input}\\
\hline
$h_{\rm HVC}$    & scale height of HVC population  & $6.67$ kpc \\
$f_{\rm v,0}$    & central HVC volume fraction     & $0.0185$ \\
$\langle n_{\rm HI}\rangle$  & mean H\,{\sc i} volume density       & $0.1$ cm$^{-3}$ \\
$z_{\rm IVC}$    & upper $z$-height limit for IVCs & $5$ kpc \\
$v_{\rm infall}$ & infall velocity                 & $100$ km\,s$^{-1}$ for HVCs\\
                 &                                 & $50$ km\,s$^{-1}$ for IVCs \\
\hline
\multicolumn{3}{c}{Output}\\
\hline
$f_{\rm HVC}(r)$                    & projected HVC covering fraction for radii $r\geq9$ kpc    &
$2.1$\,exp$(-r/12)$ \\
$f_{\rm HVC}(r)$                    & projected HVC covering fraction for radii $r<9$ kpc       & 1 \\
$\langle f_{\rm IVC,MW} \rangle$    & mean IVC covering fraction from inside the sphere         &
$0.30$ \\
$\langle f_{\rm HVC,MW} \rangle$    & mean HVC covering fraction from inside the sphere         &
$0.29$ \\
$\langle f_{\rm HVC} \rangle$       & mean HVC/IVC covering fraction from an outside vantage point  &
$0.21$ \\
$r_3$                               & radius for which $\langle f_{\rm HVC} \rangle \geq 0.03$    &
50 kpc\\
$\langle N$(H\,{\sc i})$ \rangle $  & mean  H\,{\sc i} column density in HVCs/IVCs from inside the sphere &
$1.3\times 10^{19}$ cm$^{-2}$ \\
$M_{\rm HI,tot}$                    & total neutral gas mass in HVCs/IVCs at $r\leq r_3$ &
$1.2 \times 10^8\,M_{\sun}$ \\
$dM_{\rm HI,tot}/dt$                & total neutral gas mass accretion rate for gas at $r\leq r_3$ &
$0.74\,M_{\sun}\,$yr$^{-1}$\\
\enddata
\end{deluxetable*}

As shown by Braun \& Thilker (2004), the high-resolution 
21cm HVC data of M31 from Thilker et al.\,(2004) follows 
a ``standard'' H\,{\sc i} CDDF in the form 
$f(N_{\rm HI})\propto N_{\rm HI}^{-\beta}$ with $\beta\approx 1.5$
in the column density range log $N$(H\,{\sc i}$)=18-20$. 
The slope is very similar to the one derived for the 
Milky Way HVCs (Lockman et al.\,2002; previous 
subsection) and is in good agreement with values
derived for low-redshift QSO H\,{\sc i} absorption-line
systems (see Sect.\,5.1).

%

\section{Geometrical modeling of HVCs in the Local Group}

\subsection{Modeling setup}

To combine the observational information on the HVC population 
of the Milky Way and M31 we have developed the custom-written
numerical code {\tt halopath}, which is based on a simple 
geometrical model assuming spherical symmetry. The code 
allows us to model the radial distribution of gas in the
halos of galaxies, its mass distribution and its 
absorption-cross section from any given vantage point
inside and outside the sphere.
Because of the unknown size distribution of HVCs
we here do no attempt to model individual H\,{\sc i} {\it clouds} 
as HVC analogs, but instead consider the volume-filling
factor of neutral gas as main input parameter, from which 
all relevant physical quantities (e.g., projected covering fraction,
H\,{\sc i} column density, H\,{\sc i} mass) can be easily
obtained and compared to observations.

To support the model with the necessary observational data, we 
assume that the HVC populations of the Milky Way and M31 
are identical in a statistical sense (same
radial distribution and same volume filling factor of
gas with log $N$(H\,{\sc i}$)\geq 17.5$). In view of the 
similarity of both galaxies in terms of morphology,
mass, luminosity, etc. this assumption is justified.
As mentioned above, the key parameter that describes
the spatial distribution of HVCs in our spherical model is the 
radius-dependent volume-filling factor of optically
thick HVC gas with log $N$(H\,{\sc i}$)\geq 17.5$, 
$f_{\rm v,HVC}(R)$, where $R$ is the physical radius (compared 
to the projected radius $r$). Because the projected (area) 
covering fraction of HVCs in the M31 halo can be 
described by an exponential (see above), 
we assume that $f_{\rm v,HVC}(R)$ follows
an exponential, too, so that we write:
\begin{equation}
f_{\rm v,HVC}=f_{\rm v,0}\,{\rm exp}\,(-R/h_{\rm HVC}).
\end{equation}
In this equation, $f_{\rm v,0}$ is the volume-filling factor
in the center of the sphere and $h_{\rm HVC}$ is the 
scale height of the HVC population. Another important
parameter in our model is the mean H\,{\sc i} volume
density $\langle n_{\rm HI}\rangle$ in the HVC gas, 
as $\langle n_{\rm HI}\rangle$
together with $f_{\rm v,HVC}(R)$ determines the 
neutral gas mass in HVCs per radial bin (= radial volume element),
$M_{\rm HI}(r)=f_{\rm v,HVC}(R)\,\langle n_{\rm HI}\rangle\,\mu m_{\rm H}$,
with $m_{\rm H}$ as hydrogen mass and $\mu$ as a factor
that corrects for the presence of helium and heavy elements in the gas. 
In addition, $\langle n_{\rm HI}\rangle$
determines the mean H\,{\sc i} column density, $\langle 
N$(H\,{\sc i})$\rangle$, 
measured along any given line of sight through the halo,
since $\langle N$(H\,{\sc i}$)\rangle=\langle 
n_{\rm HI}\rangle\,d$, where $d$ is the
absorption-path length through the halo. The parameters
$\langle N$(H\,{\sc i})$\rangle$ and $\langle n_{\rm HI}\rangle$
are constrained by 21cm observations and ionization models
of Galactic IVCs and HVCs. Note that in this study we 
do {\it not} model the gas physics in the halo clouds, but
consider only the spatial distribution of neutral halo gas, 
its H\,{\sc i} column density distribution, its total
mass, and its infall rate.
  
To model the sky covering fraction of neutral 
gas from a vantage point {\it inside} the sphere 
(i.e., to model the projected neutral gas distribution in
the Milky Way halo from the position of the sun) we
introduce additional constraints based on results from 
21cm observations of IVCs and HVCs. First, we only
consider H\,{\sc i} gas as neutral halo gas if it is
located at vertical distances $z>600$ pc from the 
midplane of the disk (whose orientation can be chosen
in the model). Second, we separate neutral halo gas 
close to the disk (small $z$-heights) 
from more distant halo clouds via the parameter 
$z_{\rm IVC}$. This parameter enables us to 
distinguish between IVCs ($z\leq z_{\rm IVC}$) 
and HVCs ($z> z_{\rm IVC}$). Finally, the parameter 
$v_{\rm infall}$ defines the infall velocity of the
neutral halo gas; $v_{\rm infall}$ can be a function
of $R$ or $z$, or can be chosen to be constant 
for IVCs and HVCs, respectively. The neutral gas 
(mass) accretion rate per radial bin then is given by
$dM_{\rm HI}(r)/dt=M_{\rm HI}(r)\,v_{\rm infall}/r$. 

\subsection{Modeling results}

Using our geometrical model we are able reproduce the 
observed properties of the HVC population of the 
Milky Way and M31 using appropriate values for the above 
discussed input parameters. Our favorite model is 
summarized in Table 2. In this model, the scale-height 
of the HVC population is $h_{\rm HVC}=6.67^{+0.53}_{-0.41}$ kpc
and the central volume-filling factor of neutral 
halo gas is $f_{\rm v,0}=0.0185\pm0.0036$. 
The errors have been determined numerically; they reflect
the 1$\sigma$ error range of the exponential parameters
for $f_{\rm HVC}$ in M31 (Sect.\,2.2).
In our model we assume that IVCs and HVCs are separated
at a $z$-height of $z_{\rm IVC}=5$ kpc, in line with
what is known about the distribution of IVC and HVC
distances in the Milky Way halo (Wakker et al.\,2007, 2008; 
Thom et al.\,2006, 2008). 

Only these three parameters ($h_{\rm HVC}$, $f_{\rm v,0}$, and
$z_{\rm IVC}=5$) are required to reproduce the 
observed exponential decline
of the projected covering fraction of the HVC population
of M31 ($f_{\rm HVC}=2.1$\,exp$\,(-r/12)$ for 
radii $r\geq 9$ kpc) from a 
vantage point outside the sphere {\it and} the observed covering
fractions of $\sim 30$ percent for IVCs and HVCs from a 
vantage point inside the sphere (see Sect.\,2). Note that for $r<9$ kpc 
the projected covering fraction is set to unity. The characteristic
radial extent of the HVC population around both galaxies is 
$r_3=50\pm 6$ kpc. The mean projected covering fraction of
neutral IVC/HVC gas in the halo region is
$\langle f_{\rm HVC} \rangle =0.21\pm0.02$.

Observations of H\,{\sc i}, metal ions and 
molecular hydrogen suggest that the neutral hydrogen volume densities 
in IVCs and HVCs may span a large range over at least three orders of
magnitude ($10^{-2}\leq n_{\rm HI} \leq 10$ cm$^{-3}$; e.g., 
Wakker et al.\,1999, 2004; Richter et al.\,2003a, 2003b,
2009; Sembach et al.\,2001). For our model we adopt a value of 
$\langle n_{\rm HI} \rangle =0.1$ cm$^{-3}$, which is close 
to the one that has been derived for high-velocity cloud Complex C 
(Wakker et al.\,1999). We consider this value as a realistic estimate
for the volume-averaged mean neutral hydrogen density in IVCs and HVCs.
As infall velocity we adopt $v_{\rm infall}=100$ km\,s$^{-1}$ for HVCs
and $50$ km\,s$^{-1}$ for IVCs, assuming that the (vertical) 
infall velocities towards the disk are comparable to
the observed radial velocities of IVCs and HVCs.
Since the infall velocity of HVCs
is determined by the balance between the gravitational force and
the ram-pressure force provided by the surrounding hot coronal gas 
(e.g., Benjamin \& Danly 1997; Br\"uns \& Mebold 2004), 
a more precise modeling of 
$v_{\rm infall}$ would also require the modeling of the
density distribution of hot halo gas, which is beyond
the scope of this study. However, from 
grid-based hydrodynamical simulations of HVCs 
Heitsch \& Putman (2009) derive infall velocities
that are very similar to the velocities adopted by us.

With the above given values for $n_{\rm HI}$ and $v_{\rm infall}$ the
mean H\,{\sc i} column density in IVCs and HVCs (from the 
interior view) is $\langle N$(H\,{\sc i})$ \rangle = 1.3 \times 10^{19}$ 
cm$^{-2}$, the total H\,{\sc i} mass in the halo at 
$r\leq r_3$ is $M_{\rm HI,tot}=1.2 \times 10^8\,M_{\sun}$, 
and the total H\,{\sc i} mass accretion rate for gas at $r\leq r_3$
is $dM_{\rm HI,tot}/dt=0.74\,M_{\sun}\,$yr$^{-1}$. These
values are in excellent agreement with the observations
(see $\S2$).

%

\begin{figure*}[t!]
\epsscale{1.0}
\plotone{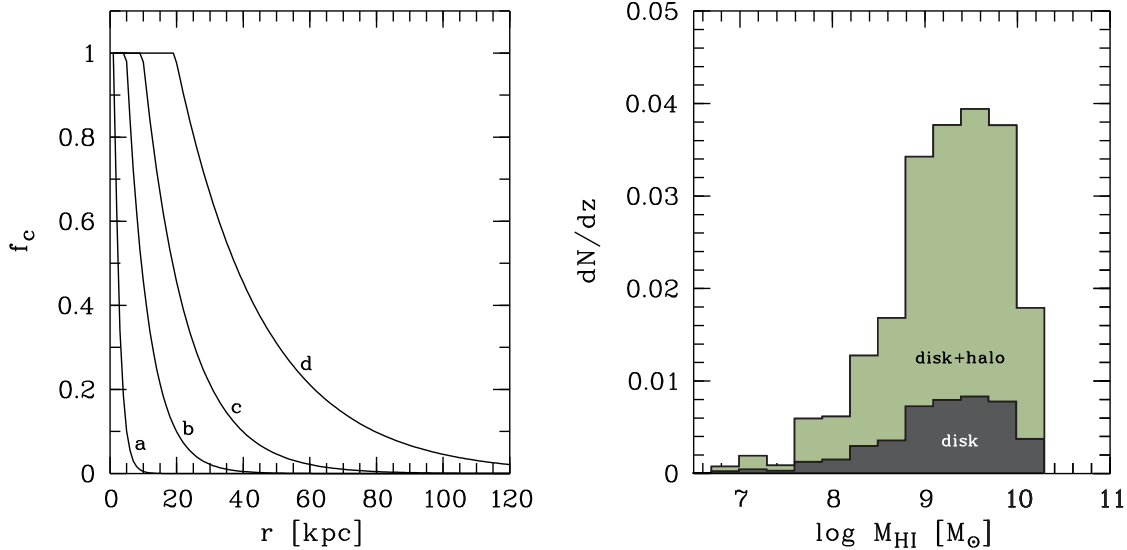}
\caption{
{\it Left:} projected covering fraction of HVCs around galaxies,
$f_{\rm HVC}$, as a function of the projected radius, $r$,
for four different H\,{\sc i} disk masses
(log $M_{\rm HI,disk}=7.75$(a)$,8.95$(b)$,9.55$(c)$,10.15$(d);
see equation 5). {\it Right:} expected number density, $d{\cal N}/dz$,
of optically thick H\,{\sc i} absorbers originating in galaxy disks
(log $N$(H\,{\sc i}$)\geq 20.3$; gray-shaded area) and disks plus HVCs
(log $N$(H\,{\sc i}$)\geq 17.5$; green-shaded area) per galaxy
H\,{\sc i} mass bin.
}
\end{figure*}

\section{Modeling of galaxy absorbers}

\subsection{Galaxies and their absorption characteristics}

In the previous section we have demonstrated that it is
possible to reproduce the statistical properties of the 
HVC population of the Milky Way and M31 using a model that is
based on very simple geometrical assumptions. In the following, 
we want to generalize our HVC model for the local galaxy population
to constrain the absorption cross section of HVCs in the local 
Universe. 

Galaxies and their circumgalactic gaseous environment can be
traced by intervening absorption lines of H\,{\sc i} and
metal ions in the spectra of distant QSOs and AGN. The 
strongest intervening neutral-gas absorbers are the so-called 
Damped Lyman $\alpha$ Absorbers (DLAs), which have H\,{\sc i} 
column densities log $N$(H\,{\sc i}$)\geq 20.3$. 
These systems contain a substantial fraction of the neutral gas mass
in the Universe (Wolfe et al.\,1995). Although there still 
is no consistent picture about the host galaxies of DLAs,
observations suggest that a mixed population of galaxies contribute
to the absorption-cross section of DLAs at $z\approx0$
(e.g., Turnshek et al.\,2001; Chen \& Lanzetta 2003; Rao et al.\,2003).
From H\,{\sc i} 21cm observations of the local galaxy population
Zwaan et al.\,(2005) concludes, however, that the total DLA cross section
at $z\approx 0$ is dominated by the gaseous disks of $L^{\star}$ and 
sub-$L^{\star}$ galaxies with H\,{\sc i} masses $>10^9\,M_{\sun}$.

Based on H\,{\sc i} 21cm observations of the Milky Way, M31,
and other nearby galaxies it is expected that neutral gas absorbers 
in the extended {\it halos} of galaxies (i.e., HVC analogs) 
have H\,{\sc i} column densities {\it below} that of DLAs
(e.g., Wakker 2004; Thilker et al.\,2004; Lockman et al.\,2002;
Murphy et al.\,1995).
Using the common absorber classification scheme,
halo absorbers therefore are expected to be
seen as so-called sub-DLAs 
($19.0\leq $\,log $N$(H\,{\sc i}$)<20.3$) and
Lyman-Limit Systems (LLS; $17.2 \leq$\,log $N$(H\,{\sc i}$)<19.0$).
A large fraction of the LLS at low redshift therefore may represent
distant analogs of the HVCs seen around the Milky Way and M31
(Richter et al.\,2011).

As for HVCs, the H\,{\sc i} column-density distribution function of 
intervening QSO absorbers at low and high redshift below
the DLA column-density limit can be 
fitted by a power-law in the form $f(N_{\rm HI})\propto N_{\rm HI}^{-\beta}$.
For QSO absorbers, $\beta$ has values between $1-2$, depending on redshift
and the column density interval chosen (see Lehner et al.\,2007). Unfortunately,
$\beta$ is poorly constrained for log $N$(H\,{\sc i}$)>16$ at $z=0$
due to the limited amount of low-redshift H\,{\sc i} absorption-line data 
in the UV. In contrast, for high $z$ there exists a large data base
that allows us to constrain $\beta$ at a relatively high accuracy 
(Ribaudo, Lehner \& Howk 2011).

The incidence of intervening DLAs, sub-DLAs and LLS in QSO spectra,
usually expressed by the quantity $d{\cal N}_{\rm LLS}/{dz}$, the number
of optically thick H\,{\sc i} absorbers per unit redshift, can be obtained
from the integration of the H\,{\sc i} CDDF over the appropriate 
column-density range (in our case log $N$(H\,{\sc i}$)>17.5$). 
Moreover, $d{\cal N}_{\rm LLS}/{dz}$ is
proportional to the space density of galaxies, 
$n_{\rm gal}$, and the mean geometrical cross section 
of optically thick H\,{\sc i}, $\langle A_{\rm HI} \rangle$, 
in these galaxies :
\begin{equation}
\frac{d{\cal N}_{\rm LLS}}{dz}=
\int_{N_{\rm HI}\geq N_{\rm LLS}}f(N_{\rm HI})\,dN_{\rm HI} =
\frac{c\,n_{\rm gal}\,\langle A_{\rm HI} \rangle}{H(z)}.
\end{equation}
We adopt
$H(z)=H_0\,(\Omega_{\rm m}\,(1+z)^3+\Omega_{\Lambda})^{1/2}$, 
$H_0=73$ km\,s$^{-1}$\,Mpc$^{-1}$, $\Omega_{\rm m}=
0.238$, and $\Omega_{\Lambda}=0.762$ (Spergel et al.\,2007).
We assume that the covering fraction of optically thick
H\,{\sc i} gas is $\langle f_{\rm HVC} \rangle =1$ in the
(inclined) disk of a galaxy and $\langle f_{\rm HVC}\rangle <1$ in the
surrounding halo, where H\,{\sc i} arises in the form 
of HVCs. Let $A_{\rm disk}$ be the geometrical cross
section of the (inclined) disk and $A_{\rm halo}=\pi R_{\rm halo}^2$
the cross section of the (spherical) halo region with
radius $R_{\rm halo}$. We then can introduce an {\it effective
HVC cross section} for an individual galaxy, 
$A_{\rm HVC,eff}=\langle f_{\rm HVC}\rangle (A_{\rm halo}-A_{\rm disk})$,
so that the total area covered by optically thick
H\,{\sc i} in and around a galaxy is 
$A_{\rm HI}=A_{\rm disk}+A_{\rm HVC,eff}$. 

From H\,{\sc i} 21cm measurements of the local
galaxy population, Zwaan et al.\,(2005) have derived 
a DLA number density per unit redshift of 
$(d{\cal N}/dz)_{\rm DLA}=0.045\pm0.006$, a result that is 
good agreement with previous estimates (e.g., Rosenberg 
\& Schneider 2003; $(d{\cal N}/dz)_{\rm DLA}=0.053\pm0.013$). 
These measurements provide direct information
on the individual geometrical cross section (i.e., $A_{\rm disk}$) 
of the gaseous disks at log $N$(H\,{\sc i}$)\geq 20.3$ in 
low-redshift galaxies for a large range of galaxy morphologies 
and luminosities. These H\,{\sc i} surveys are, however, not 
sensitive enough and do not provide sufficient spatial resolution
to identify a possibly existing HVC population in the 
halos of these galaxies. 

Based on the above-given relations,
a QSO sightline passing through both an H\,{\sc i} disk and HVC gas
of a galaxy would show a DLA as absorption signature, while a 
sightline passing only through an optically thick HVC would exhibit 
a sub-DLA or LLS. We then can write for the total number density of 
H\,{\sc i} absorbers with log $N$(H\,{\sc i}$)\geq 17.5$ 
that trace gas disks of galaxies {\it and} their surrounding 
HVC population:
\begin{equation}
\left( \frac{d{\cal N}}{dz} \right)_{\rm disk+HVC} =
\left( \frac{d{\cal N}}{dz} \right)_{\rm DLA}\,
\frac{A_{\rm disk}+A_{\rm HVC,eff}}{A_{\rm disk}}.
\end{equation}

Our goal is to estimate $A_{\rm HVC,eff}$ from a 
generalized version of our HVC model for the local 
galaxy population (for which $A_{\rm disk}$ is known).
This will enable us to estimate $(d{\cal N}/dz)_{\rm disk+HVC}$ 
and link the HVC population at $z=0$ with the 
H\,{\sc i} column-density distribution function of 
low-redshift QSO absorbers.

\subsection{On the absorption cross section of neutral gas disks}

\begin{deluxetable*}{cccccccc}
\tabletypesize{\footnotesize}
\tablewidth{0pt}
\tablecaption{Properties$^{\rm s}$ of H\,{\sc i} absorbing galaxies}
\tablehead{
\colhead{log\,$M_{\rm HI,disk}$} & \colhead{log\,$A_{\rm disk}$}    & \colhead{$r_3$} &
\colhead{$\langle f_{\rm HVC} \rangle$}          & \colhead{log\,$A_{\rm HVC,eff}$} &
\colhead{$(d{\cal N}/dz)_{\rm disk+HVC}$} & \colhead{log\,$M_{\rm HI,HVC}$} &
\colhead{$dM_{\rm HI}/dt$} \\
\colhead{$[M_{\rm HI}$ in $M_{\sun}]$} & \colhead{[$A$ in kpc$^2]$} & \colhead{[kpc]} &
\colhead{} & \colhead{$[A$ in kpc$^2]$} & \colhead{} & \colhead{$[M_{\rm HI}$ in $M_{\sun}]$} &
\colhead{$[M_{\sun}\,$yr$^{-1}]$} \\
}
\startdata
$6.7-7.0$   & 0.03 &   2 & 0.35 & 0.41 & $7.1\times 10^{-4}$ & 3.38 & $5.9\times 10^{-5}$ \\
$7.0-7.3$   & 0.33 &   3 & 0.23 & 0.88 & $1.9\times 10^{-3}$ & 4.21 & $3.7\times 10^{-4}$ \\
$7.3-7.6$   & 0.63 &   4 & 0.21 & 0.97 & $8.4\times 10^{-4}$ & 4.82 & $1.4\times 10^{-3}$ \\
$7.6-7.9$   & 0.93 &   7 & 0.15 & 1.51 & $5.9\times 10^{-3}$ & 5.46 & $5.2\times 10^{-3}$ \\
$7.9-8.2$   & 1.23 &   9 & 0.17 & 1.73 & $6.1\times 10^{-3}$ & 5.91 & $1.2\times 10^{-2}$ \\
$8.2-8.5$   & 1.53 &  13 & 0.18 & 2.05 & $1.3\times 10^{-2}$ & 6.39 & $3.1\times 10^{-2}$ \\
$8.5-8.8$   & 1.83 &  19 & 0.18 & 2.40 & $1.7\times 10^{-2}$ & 6.85 & $7.5\times 10^{-2}$ \\
$8.8-9.1$   & 2.13 &  27 & 0.19 & 2.70 & $3.4\times 10^{-2}$ & 7.29 & $0.17$ \\
$9.1-9.4$   & 2.43 &  39 & 0.19 & 3.00 & $3.8\times 10^{-2}$ & 7.73 & $0.40$ \\
$9.4-9.7$   & 2.73 &  55 & 0.20 & 3.30 & $3.9\times 10^{-2}$ & 8.15 & $0.85$ \\
$9.7-10.0$  & 3.03 &  78 & 0.20 & 3.61 & $3.8\times 10^{-2}$ & 8.56 & $1.69$ \\
$10.0-10.3$ & 3.33 & 110 & 0.22 & 3.91 & $1.8\times 10^{-2}$ & 8.95 & $3.09$ \\
\enddata
\tablenotetext{a}{{\it Explanations:}
$r_3=$\,halo radius beyond which the
projected HVC covering fraction is $\leq 3$ percent;
$A_{\rm disk}=$\,H\,{\sc i} disk area;
$\langle f_{\rm HVC} \rangle =$\,mean projected HVC covering fraction
for halo region; $A_{\rm HVC,eff}=f_{\rm HVC}(A_{\rm halo}-A_{\rm disk})$;
$(d{\cal N}/dz)_{\rm disk+HVC}=$\,number density of disk/halo absorbers
with log $N$(H\,{\sc i}$)\geq 17.5$ (per galaxy H\,{\sc i} mass bin);
$M_{\rm HI,HVC}=$\,total neutral gas mass in HVCs per galaxy;
$dM_{\rm HI}/dt=$\,neutral gas mass infall rate per galaxy.}
\end{deluxetable*}

Before we start to investigate the absorption cross-section 
of neutral gas in the halos of galaxies, it is useful to
briefly discuss the relation between the absorption cross 
section, the total H\,{\sc i} mass, and the scale-length of
neutral gas disks in local galaxies, as derived from 
H\,{\sc i} 21cm surveys.
Using 21cm data from Arecibo and the {\it Very Large Array} (VLA)
Rosenberg \& Schneider (2003) have studied in detail these and other
properties of 50 nearby galaxies in the context of the
low-redshift DLA population.
Rosenberg \& Schneider find that the total 
(inclination-corrected) H\,{\sc i} 
cross section of gas disks in their galaxy sample 
in the DLA column-density range (log $N\geq20.3$) is given by
$A_{\rm disk}=\pi\,a\,b\,/4$, where 
$a$ and $b$ are the major and minor axis parameters (in kpc)
derived at the DLA column density limit (see also
Rosenberg \& Schneider 2003, their appendix.)
They also find a remarkably tight correlation between
$A_{\rm disk}$ and the total H\,{\sc i} mass of the disk, 
$M_{\rm HI}$ (in solar mass units), as
log\,$A_{\rm disk}=\,$log\,$M_{\rm HI}-6.82$.

Since the mean value for $b$ in a sample of randomly inclined
gas disks is expected to be $0.637\,a$, we can
write for the mean disk area 
$\left< A_{\rm disk} \right> \approx 0.16\,\pi\,a^2=0.5\,a^2$ or
log\,$\left< A_{\rm disk} \right> \approx 2\,$log$\,a-0.3$.
Combining this with the observed relation between $A_{\rm disk}$ 
and $M_{\rm HI}$, we obtain a relation between the total 
H\,{\sc i} mass and the disk radius at the DLA limit in the form
log\,$a=0.5\,$log\,$M_{\rm HI}-3.26$. Finally, 
Rosenberg \& Schneider (2003) find a (weak) correlation between the 
H\,{\sc i} mass and the J-band luminosity in their sample,
log\,$M_{\rm HI}=0.44$\,log\,$L_J+5.27$, where $L_J$ is 
in solar luminosity units.

\subsection{On the absorption cross section of HVCs surrounding neutral gas disks} 

Surveys of the local galaxy population, 
such as the one presented by Rosenberg \& Schneider (2003), 
show that galaxies span several orders of magnitude 
in parameters like H\,{\sc i} mass, optical luminosity, 
and H\,{\sc i} disk size. One crucial question
that concerns the cross section of neutral gas in the 
halos of these galaxies is, how the size of the 
gaseous halo of a galaxy is related to the
above listed parameters. 

The standard approach to scale the size 
of a galaxy's gaseous halo with its luminosity is to adopt a 
Holmberg-like luminosity scaling, so that the halo
radius is given by
\begin{equation}
R_{\rm halo}(L)=R_{\star}\,\left ( \frac{L}{L^{\star}} \right)^{\delta},
\end{equation}
where $\delta\approx0.2$ and $R_{\star}\approx 110$ kpc for B-band
luminosities, as derived from analyses of 
intervening Mg\,{\sc ii} absorbers (which trace neutral
{\it and} ionized gas in disks and halos) and their relation
to galaxies (Steidel 1995; Kacprzak et al.\,2008).
It can be shown that with the above given relations between 
$L_J$, $M_{\rm HI}$, and $A_{\rm disk}$ a Holmberg-like 
luminosity scaling as given by equation (4) corresponds to 
a linear scaling of the halo radius with the H\,{\sc i} disk radius.
This is because $M_{\rm HI}\propto L^{0.44}$ 
(Rosenberg \& Schneider 2003) and
$a \propto M_{\rm HI}^{1/2}$, which leads to
$R_{\rm halo}\propto L^{0.22}\propto a$, i.e., 
a Holmberg-like luminosity scaling with $\delta=0.22$.
For the following, we therefore assume a Holmberg-like 
luminosity scaling with $\delta=0.22$,
but we parametrize the halo size over the relation
$R_{\rm halo}(a)=\gamma\,a$, where $\gamma>1$
and $a$ is the H\,{\sc i} disk radius for
H\,{\sc i} column densities above the DLA limit.
For M31 $\gamma\approx 1.7$ (see Sect.\,2.2).

To characterize the covering fraction of HVCs, $f_{\rm HVC}$, 
in galaxy halos, we assume that the exponential decline
of $f_{\rm HVC}$ observed in M31 reflects a general 
behavior of HVCs in galaxies in the local Universe.
We then can express the projected 
covering fraction of HVCs around galaxies as a function 
of its H\,{\sc i} disk length in the form
\begin{equation}
f_{\rm HVC}(r) \approx \left\{ \begin{array}{l@{\quad:\quad}l}
  1 & r\leq0.3\,a\\
  2.1\cdot\,{\rm exp\,}(-2.5\,r/a) & r>0.3\,a\\
  \end{array}
  \right\} .
\end{equation}
For M31, $a=30$ kpc (Braun et al.\,2009; see above).
In Fig.\,2, left panel, we show $f_{\rm HVC}(r)$ for 
four different H\,{\sc i} masses.
In analogy to what has been discussed for the M31
HVC population, we define $r_3$ as the halo radius,
beyond which $f_{\rm HVC}$ falls below the 3-percent level
and set $R_{\rm halo}= r_3$. The scaling relation
between the disk and halo radius in our model then comes out to
$R_{\rm halo}=1.69\,a$ for all halo radii considered (i.e., $\gamma=1.69$).

Since log\,$a=0.5\,$log\,$M_{\rm HI}-3.26$ (see above), 
equation (5) allows us to calculate the sizes of 
neutral gas halos of low-redshift galaxies as a function of 
their H\,{\sc i} mass and luminosity. It also enables us
to predict the cross section and number density of sub-DLAs
and LLS that represent HVC analogs, their radial distribution,
and estimate the neutral-gas accretion 
rate of galaxies in the local Universe.

\subsection{Modeling results}

As input for our generalized HVC model we adopt the 
H\,{\sc i} mass distribution of low-redshift DLAs
from Zwaan et al.\,(2005), as derived from 
a high-resolution 21cm survey of the local galaxy population.
Using 21cm data from the WRST Zwaan et al.\,(2005) have studied 
the H\,{\sc i} properties of 355 nearby galaxies and their
contribution to the local DLA population. The expected 
values of $(d{\cal N}/dz)_{\rm DLA}$ for the different
H\,{\sc i} masses of the galaxies in their sample
are shown as gray-shaded area in the right panel 
of Fig.\,2. Obviously, galaxies with H\,{\sc i} masses
in the range log\,$M_{\rm HI}=8.8-10.0$ dominate the
absorption cross section of DLAs at $z=0$.

Based on these data, we derive for each 
galaxy H\,{\sc i} mass bin the radius and area of the
H\,{\sc i} disk and the H\,{\sc i} halo using
the above discussed relations.
Note that we here do not take into account
the possibility that the neutral gas disks of galaxies
significantly extend below the DLA column-density limit.
Using our generalized HVC model we then calculate 
for each mass bin the radius-dependent (projected) 
HVC covering fraction 
($f_{\rm HVC}$; equation 5), the effective HVC cross 
section ($A_{\rm HVC,eff}$), and the expected number 
density of disk+halo H\,{\sc i} absorbers (equation 3).
Moreover, our model calculates for each galaxy H\,{\sc i} 
mass bin the mean projected HVC covering fraction
($\langle f_{\rm HVC} \rangle$), the total neutral gas
mass in HVCs ($M_{\rm HI,HVC}$), and the neutral gas mass 
accretion rate ($dM_{\rm HI}/dt$). For the
infall velocities we adopt the values for 
$v_{\rm infall}$ for IVCs and HVCs listed in Table 2.
The expected number densities
$(d{\cal N}/dz)_{\rm disk+HVC}$ as a function of the
galaxy H\,{\sc i} mass are indicated with the green-shaded 
area in the right panel of Fig.\,2. All results are summarized
in Table 3. Based on these results we
derive the following relation between the H\,{\sc i} disk
mass of galaxies, log\,$M_{\rm HI}$ (in solar units), and the radius 
of the neutral gas halo, $R_{\rm halo}$ (in [kpc]):
\begin{equation}
{\rm log}\,R_{\rm halo}=0.5\,{\rm log}\,M_{\rm HI} - 3.03.
\end{equation}
If we integrate the values of $(d{\cal N}/dz)_{\rm disk+HVC}$
listed in Table 3 over the entire mass range, we derive a total
number density of disk/halo absorbers of
$(d{\cal N}/dz)_{\rm disk+HVC}=0.212$. This value is $\sim 5$ times
larger than the number density of DLAs at $z=0$ (Zwaan et al.\,2005),
suggesting that the absorption-cross section of HVCs with 
log $N$(H\,{\sc i}$)\geq 17.5$ exceeds that of DLAs 
by a factor of $\sim 4$, on average.
As for DLAs, the total absorption cross section of optically thick
H\,{\sc i} in disks and halos is dominated by galaxies in the mass range
log\,$M_{\rm HI}=8.8-10.0$.
As our model indicates, the mean (projected) covering
fraction of HVCs in galaxy halos is small, 
$\langle f_{\rm HVC} \rangle=0.2$, typically.

\begin{figure*}[t!]
\epsscale{1.0}
\plotone{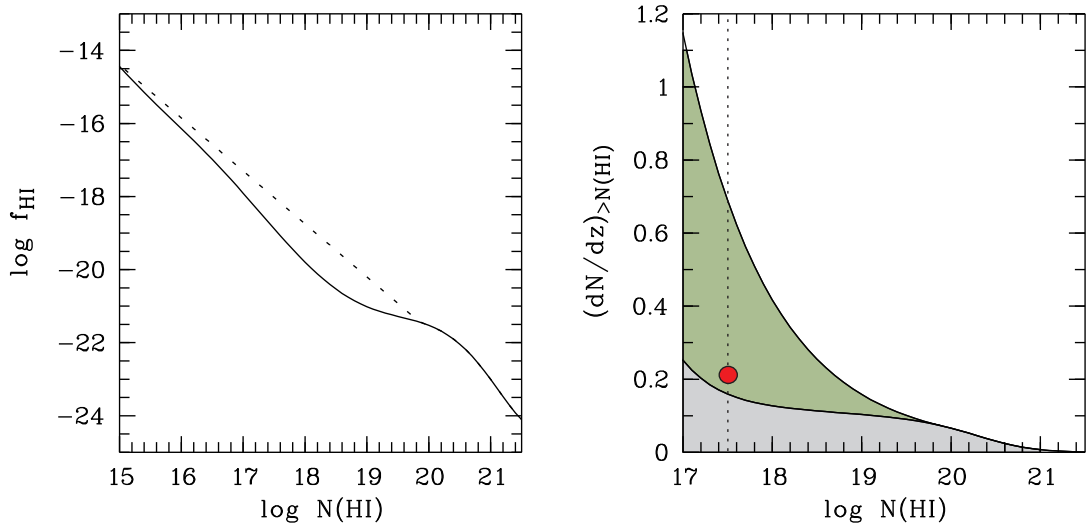}
\caption{
{\it Left:} H\,{\sc i} column-density distribution function (CDDF)
at $z\approx0$, based on results by Corbelli \& Bandiera (2002; solid line),
Zwaan et al.\,(2005; solid line) and Lehner et al.\,(2007; dashed
line). {\it Right:} cumulative number density of H\,{\sc i} absorbers
per unit redshift ($(d{\cal N}/dz)_{>N({\rm HI})})$
as a function of the cutoff H\,{\sc i} column density.
The gray-shaded area indicates
the allowed $d{\cal N}/dz$ range for the CDDF
based on the data of Corbelli \& Bandiera (2002) and Zwaan et
al.\,(2005); the gray-shaded area
plus the green-shaded area shows the allowed $d{\cal N}/dz$
range for the CDDF extrapolated from
the Lehner et al.\,(2007) data.
The red filled circle indicates the value for
$(d{\cal N}/dz)_{\rm disk+HVC}$ derived from our HVC model.
}
\end{figure*}

\subsection{Neutral gas accretion-rate density}

Using the mass accretion rates ($dM_{\rm HI}/dt$) listed
in Table 3 together with the H\,{\sc i} mass function of the local
galaxy population (Zwaan et al.\,2005) we can 
estimate the neutral gas accretion-rate density of HVCs
(mass accretion rate per unit volume) at low redshift.
We obtain $dM_{\rm HI}/dt/dV = 0.022\,M_{\sun}$\,yr$^{-1}$\,
Mpc$^{-3}$. Note that this value is calculated under
the assumption that all of the neutral gas in the
halos of galaxies is being accreted onto their disks,
independently of its origin inside or outside the
host galaxies. The above estimate does not include the
mass of the ionized gas component of HVCs, which may
contribute substantially to the total mass of multi-phase
halo clouds (e.g., Fox et al.\,2010; Winkel et al.\,2011).
Also not included are partly neutral
gas fragments with masses and angular sizes below the
detection limit of 21cm HVC surveys. Such structures 
are known to exist in the Milky Way halo (Richter et al.\,2009),
but their (total) neutral gas mass most likely is small compared
to the large, extended 21cm HVCs. The role of ionized gas 
is further discussed below.

The value of $0.022\,M_{\sun}$\,yr$^{-1}$\,
Mpc$^{-3}$ is remarkably close to the star-formation rate
density at $z=0$ ($\dot{\rho}_{\star}=0.01-0.02\,M_{\sun}$\,yr$^{-1}$\,
Mpc$^{-3}$, as derived from ultraviolet and infrared 
observational data (Hopkins \& Beacom 2006). Therefore,
cold-gas accretion by HVCs possibly plays an important
(if not dominating) role in feeding galaxies at $z\approx0$
with gaseous material to power star formation.

The above given value for the accretion-rate density
can also be compared with recent estimates of the ``cold-mode''
gas accretion-rate densities at $z=0$ from cosmological simulations.
Using an SPH code, Kere$\check{s}$ et al.\,(2009) find 
$dM/dt/dV \sim 0.03\,M_{\sun}$\,yr$^{-1}$\,Mpc$^{-3}$
for cold gas that never exceeded a maximum temperature
of $T_{\rm max}=2.5\times10^5$ K.
However, using SPH simulations with more realistic gas physics
van\,de\,Voort et al.\,(2011) derive a much lower 
``cold-mode'' gas accretion-rate density for 
{\it galaxies} at $z=0$ of $dM/dt/dV \sim 0.002\,M_{\sun}$\,
yr$^{-1}$\,Mpc$^{-3}$, while for the superordinate
DM halos the cold-mode accretion-rate 
is estimated to be $dM/dt/dV \sim 0.01\,M_{\sun}$\,
yr$^{-1}$\,Mpc$^{-3}$, thus five-times higher.
Therefore, only 20 percent of the
cold gas that enters the DM halo in their simulation is 
actually being accreted as cold gas by the central 
galaxy. Unfortunately, these 
studies do not provide information on the absorption cross section
of all the cold gas in the DM halos and its radial distribution
around the galaxies (independently of whether it is being accreted or not), 
so that a detailed comparison between the simulation results 
and our HVC model is not possible at this point. 

It needs to be mentioned that gas that is 
considered as ``cold'' in the cosmological simulations does
{\it not} necessarily end up as H\,{\sc i} high-velocity
gas that is detectable via 21cm observations.
It is expected that a substantial fraction of the
accreted gas that never was heated up to the virial temperature
of the host halo remains diffuse and ``warm'', i.e., at
low densities ($n_{\rm H}<10^{-2}$ cm$^{-3}$) and 
intermediate temperatures ($T=10^4-10^5$ K).
The neutral gas fraction in such warm gas is expected to
be low, so that it remains unseen in H\,{\sc i} 21cm emission
(``warm`` mode of gas accretion; Heitsch \& Putman 2009;
Bland-Hawthorn 2008). The existence of such a warm, ionized gas
component in the halos of galaxies is strongly
supported by the detection of intermediate- and
high-ion absorption (e.g., from Si\,{\sc iii},
C\,{\sc iii}, C\,{\sc iv}, and Si\,{\sc iv}) in
the halo of the Milky Way (e.g., 
Fox et al.\,2006; Sembach et al.\,1995, 1999)
and in the circumgalactic environments of 
other galaxies (e.g., Ribaudo et al.\,2011). Also the
ionized envelopes of 21cm HVC complexes represent 
significant gas reservoirs that need to be considered
for a realistic estimate of the {\it total} (neutral and
ionized) gas mass that is being accreted by galaxies 
(e.g., Winkel et al.\,2011; Fox et al.\,2010). The 
contribution of the ionized gas component to the
total gas infall rate is difficult to determine, however, as 
the infall velocity of ionized gas may be substantially
lower than for neutral gas because of hydrodynamical
effects that affect the ionized cloud envelopes, such as
gas stripping, turbulent mixing, and heat conduction. 
The same processes also affect the total cloud mass 
and the neutral gas fraction in HVCs and thus influence
the H\,{\sc i} volume filling factor in these clouds.
These aspects clearly are best resolved through high-resolution 
hydrodynamical simulations (e.g., Heitsch \& Putman 2009;
Vieser \& Hensler 2007).

%

\section{HVCs and their relation to intervening QSO absorbers}

\subsection{HVCs and H\,{\sc i} Ly\,$\alpha$ absorbers}

To estimate the contribution of H\,{\sc i} HVCs to the 
number density of optically thick H\,{\sc i} Ly\,$\alpha$
absorbers at low $z$, it is necessary to know the H\,{\sc i}
CDDF for log $N$(H\,{\sc i}$)>17.5$ at low redshift (equation 2).
As mentioned above, the CDDF at $z=0$ is poorly constrained for this
column density range owing to the fact that high-column density
H\,{\sc i} absorbers are rare and that the amount of the
QSO absorption-line data in the ultraviolet is limited. 
Corbelli \& Bandiera (2002) have combined absorption-line data 
for intermediate redshifts from Bandiera \& Corbelli (2001),
low-redshift absorption data from Weymann et al.\,(1998),
and 21cm H\,{\sc i} emission-line data from Ryan-Weber 
et al.\,(2003) to construct the H\,{\sc i} CDDF in the 
range log $N$(H\,{\sc i}$) \approx 13-21$. The CDDF presented in 
Corbelli \& Bandiera (2002) can be fitted by power laws
in the ranges log $N$(H\,{\sc i}$)<18$ and log $N$(H\,{\sc i}$)>20$,
while at log $N$(H\,{\sc i}$)=18-20$ there seems to be a 
plateau in $f(N_{\rm HI})$. Since this is exactly the column-density
range of relevance for HVCs, these data imply that 
the total absorption-cross section of HVCs increases only
mildly with a decreasing H\,{\sc i} cutoff column density.
To check the validity of this interesting feature in
$f(N_{\rm HI})$ more absorption-line data 
for the range log $N$(H\,{\sc i}$)=18-20$ are required.
To compare the frequency of sub-DLAs and LLS at low redshift with 
the HVC absorption cross section obtained from our model
we have combined the absorption/emission-line data 
from Corbelli \& Bandiera (2002) and Zwaan et al.\,(2005)
and have constructed a ``hybrid'' H\,{\sc i} CDDF 
at $z\approx 0$, in Fig.\,3, left panel, shown as 
solid line.

Independent constraints for $f(N_{\rm HI})$ at low redshift
come from the study by Lehner et al.\,(2007), how 
have analyzed low-redshift Ly\,$\alpha$ absorbers
based on high-resolution HST/{\it STIS} data.
While their study is limited to the column density 
range log $N$(H\,{\sc i}$) \approx 13-16.5$, it is 
possible to extrapolate their CDDF from the three
data points between log $N$(H\,{\sc i}$)=14.5-16.5$
(Lehner et al.\,2007; their Fig.\,14) to 
log $N$(H\,{\sc i}$)=20$ with a power law
$f(N)\propto N^{-\beta}$, where $\beta \approx 1.3$ 
(Fig.\,3, left panel, dashed line). This
extrapolation nicely connects to the 21cm
data from Zwaan et al.\,(2005), but 
lies substantially above the H\,{\sc i} CDDF
based on the Corbelli et al. results. For the 
following, we consider both representations
of the H\,{\sc i} CDDF shown in Fig.\,3 as 
plausible input parameters to estimate 
the absorption cross section
of H\,{\sc i} in sub-DLAs and LLS at $z\approx 0$.
In a recent study, Ribaudo, Lehner \& Howk (2011) have compiled
a large sample of LLS, most of them located at high redshifts 
($z\leq 2.6$). Interestingly, their data points for $f(N)$ 
fit well to the H\,{\sc i} CDDF extrapolated from
the Lehner et al.\,(2007) data.

Fig.\,3, right panel, shows the (cumulative) 
number density of H\,{\sc i} absorbers 
per unit redshift, $(d{\cal N}/dz)_{>N({\rm HI})}$,
as a function of the cutoff H\,{\sc i} column density,
$N$(H\,{\sc i}), dervied from integrating the two 
representations of the H\,{\sc i} CDDF over all
column densities larger than $N$(H\,{\sc i})\,
(equation 2). This plot now can be directly 
compared to the results from our HVC model.
The gray-shaded area indicates
the allowed $d{\cal N}/dz$ range ($0-100$ percent 
contribution of HVCs to the H\,{\sc i} absorber 
population) for the hybrid H\,{\sc i} CDDF
(Corbelli \& Bandiera 2002; Zwaan et al.\,2005;
left panel, solid line), while the gray-shaded 
plus green-shaded area shows the allowed $d{\cal N}/dz$ 
range based on the H\,{\sc i} CDDF extrapolated from
the Lehner et al.\,(2007) data (left panel, dashed line).
The red filled circle indicates the value for 
$(d{\cal N}/dz)_{\rm disk+HVC}$ derived from our HVC model.
Thus, our HVC model predicts a value for $d{\cal N}/dz$
from H\,{\sc i} disk and halo absorbers that lies 
slightly above the value expected from the 
hybrid CDDF, but lies at a $\sim 30$ percent level
of $d{\cal N}/dz$ predicted by the CDDF extrapolated from
the Lehner et al.\,(2007) data. We conclude
that HVCs in our model contribute with $30-100$ percent
to the population of H\,{\sc i} Ly\,$\alpha$ absorbers with
log $N$(H\,{\sc i}$)\geq 17.5$ at $z\approx0$. Most likely,
the contribution of HVCs is clearly less than $100$ percent,
as extended neutral and partly-ionized gas disks with
log $N$(H\,{\sc i}$)< 20.3$ and galaxy outflows (as traced
by strong Mg\,{\sc ii} absorption; see next subsection) 
contribute to the population of Ly\,$\alpha$ absorbers 
in the LLS and sub-DLA column density range.

A more detailed comparison between the absorption cross
section of HVCs and the local H\,{\sc i} Ly\,$\alpha$ 
absorber population has to await a more precise determination 
of the H\,{\sc i} CDDF at $z\approx0$ from HST/{\it COS} data.

\subsection{HVCs and intervening Mg\,{\sc ii} absorbers}

The Mg\,{\sc ii} resonant doublet near 2800 \AA\,
is commonly used to study neutral and ionized gas 
in the outskirts of galaxies at $0.3<z<2.2$
(e.g., Bergeron \& Boiss\'e 1991; Charlton \& Churchill 1998; 
Ding et al.\,2005).
The so-called ``strong'' Mg\,{\sc ii} systems represent
intervening metal absorbers in QSO spectra that have
rest-frame equivalent widths $W_{\rm} > 0.3$ \AA\, in
the Mg\,{\sc ii} $\lambda 2796$ line. They are 
commonly found within $35 h^{-1}$ kpc of luminous galaxies 
and thus most likely are related to neutral and ionized gas
in the disks and halos of low-redshift galaxies.
The number density of strong intervening Mg\,{\sc ii} 
systems in the local Universe is expected to be 
$(d{\cal N}/dz)_{\rm MgII}\approx 0.5$, as estimated from
extrapolating the redshift evolution of strong
Mg\,{\sc ii} systems in the SDSS data from $z>0.3$
down to $z\approx0$ (Nestor, Turnshek \& Rao 2005).

The value for $(d{\cal N}/dz)_{\rm MgII}$ is $\sim 2-3$ times
higher than $(d{\cal N}/dz)_{\rm disk+HVC}$ estimated from
our HVC model, suggesting a substantially larger absorption
cross section of strong Mg\,{\sc ii} systems
compared to HVC H\,{\sc i} absorbers with log 
$N$(H\,{\sc i}$)\geq 17.5$. This is expected, because with 
an ionization potential of $\sim 15$ eV Mg\,{\sc ii} 
traces both neutral {\it and} ionized gas in the disks and 
halos of galaxies.
Kacprzak et al.\,(2008) estimated a Mg\,{\sc ii} covering 
fraction of $\langle f_{\rm MgII} \rangle 
\approx 0.5$ for galaxies and their gaseous halos 
from a sample of 37 Mg\,{\sc ii} selected galaxies
at intermediate redshift. This is roughly two times
the mean value for 
$\langle f_{\rm disk+HVC} \rangle$ in our HVC model.

In the Milky Way halo, the Mg\,{\sc ii} absorption cross section
in HVCs has not been determined yet, mostly because of 
the lack of appropriate high-resolution NUV data.
The covering fractions of other low and intermediate ions 
with strong transitions in the FUV 
and with ionization potentials comparable to that of Mg\,{\sc ii} 
(e.g., C\,{\sc ii}, Si\,{\sc ii}) are larger than the covering fraction
of optically thick H\,{\sc i} (Richter et al.\,2009), but not
large enough to explain the high value for $(d{\cal N}/dz)_{\rm MgII}$
in the local Universe. Recent studies suggest, indeed, that 
many strong Mg\,{\sc ii} absorbers at low redshift arise in bipolar 
outflows and galactic winds (e.g., Bond et al.\,2001; 
Bouch\'e et al.\,2011) and thus are related to halo environments with larger 
Mg\,{\sc ii} covering fractions in galaxies that are
more actively star-forming than the Milky Way and M31. 
From our estimate for $(d{\cal N}/dz)_{\rm disk+HVC}$ we conclude 
that the contribution of infalling gas clouds (HVCs) to the 
absorption cross section of strong Mg\,{\sc ii} absorbers most likely is 
small, but not negligible ($<35$ percent).

\subsection{HVCs and intervening Ca\,{\sc ii} absorbers}

Next to the Mg\,{\sc ii} doublet in the near-UV, the 
two Ca\,{\sc ii} H\&K lines in the optical near 4000 \AA\, 
represent valuable tracers for neutral gas in the inner and
outer regions of galaxies. For instance, Ca\,{\sc ii}
absorption is frequently observed in Galactic HVCs 
(e.g., Richter, Westmeier \& Br\"uns 2005; Ben Bekhti et al.\,2008)
and is used as diagnostic line to derive distance
brackets for IVCs and HVCs from optical data
(e.g., Wakker et al.\,2007, 2008; Thom et al.\,2006, 2008).
Based on archival high-resolution optical spectra of 
more than 300 QSOs, obtained with the VLT/UVES spectrograph
Richter et al.\,(2011) have investigated 
intervening Ca\,{\sc ii} absorbers at $z\leq 0.5$ and their
relation to disk and halo gas components. They derive 
a number density per unit redshift of 
$(d{\cal N}/dz)_{\rm CaII}=0.117\pm0.044$ for Ca\,{\sc ii}
systems with log $N$(Ca\,{\sc ii}$)\geq 11.65$, which is
roughly 55 percent of the cross section of H\,{\sc i} HVCs 
with log $N$(H\,{\sc i}$)\geq 17.5$ estimated from our model.

In the Milky Way halo, the covering fraction of Ca\,{\sc ii}
for log $N$(Ca\,{\sc ii}$)\geq 11.65$ is $\sim 0.2$ 
(Ben Bekhti et al.\,2008), thus $\sim 67$ percent of that
of 21cm HVCs with log $N$(H\,{\sc i}$)\geq 17.5$.
As discussed in Richter et al.\,(2011), 
Ca\,{\sc ii} absorbers with log $N$(Ca\,{\sc ii}$)\geq 11.65$ 
predominantly trace neutral gas clouds 
with log $N$(H\,{\sc i}$)\geq 18.5$. This is due to
the presence of dust in the gas (Ca is strongly depleted
into dust grains) and the mostly sub-solar metallicities of 
the absorbers. From these numbers it follows that 
intervening Ca\,{\sc ii} absorbers arise only in specific
regions in HVCs, where the H\,{\sc i} column density 
is large enough to compensate for the Ca dust-depletion
and abundance effects.

%

\section{Conclusions}

Several previous studies have discussed 
a possible link between intervening metal-line and H\,{\sc i} 
absorbers in QSO spectra and the HVC phenomenon
in the Milky Way (e.g., Blitz et al.\,1999; 
Charlton, Churchill \& Rigby 2000;
Mshar et al.\,2007; Narayanan et al.\,2008; 
Richter et al.\,2009; Stocke, Keeney \& Danforth 2010;
Ribaudo et al.\,2011).
In this study, we move this idea forward by 
bringing together in a quantitative manner the properties 
of HVCs in the Milky Way and M31, QSO absorption-line 
statistics, and neutral gas-accretion rates of galaxies
in the local Universe. We demonstrate that it is possible
to explain the star-formation rate density in the
local Universe through the infall of cold gas in the 
form of HVCs by using a simple geometrical model. In
this model we project the observed statistical properties of the
H\,{\sc i} HVCs in the Local Group onto the local galaxy 
population, assuming a Holmberg-like luminosity scaling 
of the halo size. The main results from our modeling are 
summarized in Tables 2 and 3. 

We emphasize at this point that we have {\it not} tuned 
any of the input parameters to fit the HVC cross section 
to the H\,{\sc i} absorber population or any other 
above discussed QSO absorber observables. Our 
generalized HVC model is based solely on 
observed properties of the HVC population in
the Local Group and the local galaxy population 
without any further assumptions.
Note that the exact radial dependence of projected 
HVC covering fraction, $f_{\rm HVC}(r)$ (here assumed 
to be an exponential, based on the M31 data), is unimportant 
for the total absorption cross section of HVCs, as only 
the effective area $A_{\rm HVC,eff}$ (i.e., the product
of total halo area and the {\it mean} HVC covering 
fraction) is relevant for $(d{\cal N}/dz)_{\rm HVC}$
in a sample of randomly distributed QSO sightlines.
However, because the shape of $f_{\rm HVC}(r)$ reflects
the volume-filling factor and thus the mass distribution 
of neutral gas around galaxies, $f_{\rm HVC}(r)$ is
closely related to the H\,{\sc i} accretion rate
of galaxies $dM_{\rm HI}(r)/dt= M_{\rm HI}(r)\,v_{\rm infall}/r$.
Thus, for a more precise estimate of the neutral-gas 
accretion rate of galaxies at $z=0$ it will be 
important to constrain the radial distribution of 
HVCs around other low-redshift galaxies beyond the
Local Group using sensitive, high-resolution 21cm
observations.
Such observations need to be accompanied
by a more precise estimate of the dynamics of
neutral gas structures along their infall path through 
the hot coronal gas in the halos of galaxies of
different masses, e.g., from hydrodynamical simulations 
(Kaufmann et al.\,2009; Kere$\check{s}$ et al.\,2010).

Since most of the gaseous material that is being 
accreted by galaxies at $z=0$ may be diffuse, ionized
gas (``warm accretion''; Bland-Hawthorn 2008) rather
than cold, neutral gas in the form of 21cm HVCs, it
will be important for future studies to investigate 
the distribution and mass of ionized gas in galaxy 
halos. Diffuse ionized gas structures in the extended 
halos of galaxies are expected to have temperatures 
$T<3\times 10^5$ K, relatively low gas densities, 
and low neutral gas fractions (i.e., they 
remain unseen in 21cm HVC surveys). Such structures can 
be observed best in absorption in the FUV in the lines 
of low and intermediate ions such as C\,{\sc ii}, 
C\,{\sc iii}, Si\,{\sc ii}, and Si\,{\sc iii} 
(Fox et al.\,2006; Ribaudo et al.\,2011). 
Over the next few years large amounts of such spectral
data hopefully will become available from the many 
ongoing observational campaigns with HST/{\it COS}. Results from 
these observations can be easily implemented in our halo
model to predict the ionized-gas accretion rate
for low-redshift galaxies. 

In conclusion, the increasing amount of information on
the distribution and physical properties of gas in the
inner and outer halos of galaxies from observations, simulations,
and semi-analytic models now can be used to
substantially improve our understanding of gas-accretion
processes of galaxies in the local Universe. 
Our study has shown one possible way of how 
to combine information from the local HVC population and
QSO absorption-line systems at $z=0$ to investigate
these processes for the local galaxy population.
More detailed studies of this kind (e.g.,
including constraints for ionized halo gas) thus
could be of great importance to constrain the role
of gas-accretion processes for the ongoing formation and evolution
of galaxies at low $z$ and to characterize their connection
to the cosmic web.


%

\acknowledgments

The author would like to thank Jane Charlton, Glenn Kacprzak, and
Bart Wakker for helpful comments and interesting discussions.

%

\section*{REFERENCES}
\begin{small}

\noindent
Bandiera, R. \& Corbelli, E. 2001, ApJ, 552, 386
\noindent
\\
Ben Bekhti, N., Richter, P., Westmeier, T., \& Murphy, M.T. 2008,
A\&A, 487, 583
\noindent
\\
Benjamin, R.A. \& Danly, L. 1996, ApJ, 481, 764
\noindent
\\
Bergeron, J. \& Boiss\'e, P. 1991, A\&A, 243, 344
\noindent
\\
Birnboim, Y. \& Dekel, A. 2003, MNRAS, 345, 349
\noindent
\\
Bland-Hawthorn, J. 2008, in IAU Symposium 254,
The Galaxy Disk in Cosmological Context, 
astro-ph/08112467
\noindent
\\
Blitz, L., Spergel, D.N., Teuben, P.J., Hartmann, D., 
\& Burton, W.B. 1999, ApJ, 514, 818
\noindent
\\
Bouch\'e, N., Hohensee, W., Vargas, R., Kacprzak, G.G., Martin, C.L., 
Cooke, J., Churchill, C.W. 2011, astro-ph/11105877
\noindent
\\
Bond, N.A., Churchill, C.W., Charlton, J.C. \& Vogt, S.S. 2001,
ApJ, 557, 761
\noindent
\\
Braun, R., Thilker, D.A., Walterbros, R.A.M., \& Corbelli, E. 2009,
ApJ, 695, 937
\noindent
\\
Braun, R., \& Thilker, D.A. 2004, A\&A, 417, 421
\noindent
\\
Br\"uns, C. \& Mebold, U. 2004, in ASSL, Vol.\,312, 
High-Velocity Clouds, ed. van\,Woerden et al. 
(Kluwer Academic Publishers), 251
\noindent
\\
Charlton, J.J., Churchil. C.W., \& Rigby, J.R. 2000, ApJ, 544, 702
\noindent
\\
Charlton, J.C. \& Churchill C.W. 1998, ApJ, 499, 181
\noindent
\\
Chen, H.-W. \& Lanzetta, K.M. 2003, ApJ, 597, 706
\noindent
\\
Corbelli, E. \& Bandiera, R. 2002, ApJ, 567, 712
\noindent
\\
Ding, J., Charlton, J.C., Churchill C.W. 2005, ApJ, 621, 615
\noindent
\\
Fox, A.J., Wakker, B.P., Smoker, J.V., Richter, P., Savage, B.D.,
\& Sembach, K.R. 2010, ApJ, 718, 1046
\noindent
\\
Fox, A.J., Savage, B.D., \& Wakker, B.P. 2006, ApJS, 165, 229
\noindent
\\
Fox, A.J., Savage, B.D., Wakker, B.P., Richter, P., Sembach, K.R.,
\& Tripp, T.M. 2004, ApJ, 602, 738
\noindent
\\
Gardiner, L.T. \& Noguchi, M. 1996, MNRAS, 278, 191 
\noindent
\\
Heitsch, F. \& Putman, M.E. 2009, ApJ, 698, 1485
\noindent
\\
Hopkins, A.M. \& Beacom, J.F. 2006, ApJ, 651, 142
\noindent
\\
Kacprzak, G.G., Churchill, C.W., Steidel, C.C., \& Murphy, M.T. 2008,
AJ, 135, 922
\noindent
\\
Kaufmann, T., Bullock, J.S., Maller, A.H., Fang, T., \& Wadsley, J. 2009,
MNRAS, 396, 191
\noindent
\\
Kere$\check{s}$, D., Katz, N., Fardal, M., Dav\'e, R., \& Weinberg, D.H. 2009a,
MNRAS, 395, 160
\noindent
\\
Kere$\check{s}$, D. \& Hernquist, L. 2009b, ApJ, 700, L1
\noindent
\\
Kere$\check{s}$, D., Katz, N., Weinberg, D.H., \& Dav\'e, R. 2005, MNRAS, 363, 2
\noindent
\\
Lehner, N., Savage, B.D., Richter, P., Sembach, K.R., Tripp, T.M.,
\& Wakker, B.P. 2007, ApJ, 658, 680
\noindent
\\
Lockman, F.J., Murphy, E.M., Petty-Powell, S., \& Urick, V.J. 2002, ApJS, 140, 331
\noindent
\\
McConnachie, A.W., Irwin, M.J., Ferguson, A.M.N., Ibata, R.A., 
Lewis, G.F., \& Tanvir, N. 2005, MNRAS, 356, 979
\noindent
\\
Murphy, E.M., Lockman, F.J., \& Savage, B.D. 1995, ApJ, 447, 642
\noindent
\\
Mshar, A.C., Charlton, J.C., Lynch, R.S., Churchill, C.W.,\& Kim, T.-S. 2007,
ApJ, 669, 135
\noindent
\\
Narayanan, A., Charlton, J.C., Misawa, T., Green, R.E., \& Kim, T.-S. 2008,
ApJ, 689, 782
\noindent
\\
Nestor, D.B., Turnshek, D.A., \& Rao 2005, ApJ, 628, 637
\noindent
\\
Oosterloo, T., Fraternali, F., \& Sancisi, R. 2007, ApJ, 134, 1019
\noindent
\\
Pisano, D.J., Barnes, D.G., Gibson, B.K., Staveley-Smith, L., Freeman, K.C., 
Kilborn, V.A. 2004, ApJ, 610, L17
\noindent
\\
Rao, S.M., Nestor, D.B., Turnshek, D.A., Lane, W.M., 
Monier, E.M., \& Bergeron, J. 2003, ApJ, 595, 94
\noindent
\\
Rees, M.J., Ostriker, J.P., 1977, MNRAS, 179, 541
\noindent
\\
Ribaudo, J., Lehner, N., Howk, J.C., Werk, J., Tripp, T.M.,
Prochaska, J.X., Meiring, J.D. \& Tumlinson J. 2011, 
ApJ, 743, 207
\noindent
\\
Ribaudo, J., Lehner, N., \& Howk, J.C. 2011, ApJ, 736, 42
\noindent
\\
Richter, P., Krause, F., Fechner, C., Charlton, J.C., 
\& Murphy, M.T. 2011, A\&A, 528, A12
\noindent
\\
Richter, P., Charlton, J.C., Fangano, A.P.M., Ben Bekhti, N.,
\& Masiero, J.R. 2009, ApJ, 695, 1631
\noindent
\\
Richter, P. 2006, Reviews in Modern Astronomy, 19, 31
\noindent
\\
Richter, P., Westmeier, T., \& Br\"uns 2005,  A\&A, 442, L49
\noindent
\\
Richter, P., Sembach, K.R. \& Howk, J.C. 2003a, A\&A, 405, 1013
\noindent
\\
Richter P., Wakker B.P., Savage B.D., \& Sembach K.R 2003b, ApJ, 586, 230
\noindent
\\
Richter, P., Savage, B.D., Wakker, B.P., Sembach, K.R., Kalberla, P.M.W.
2001, ApJ, 549, 281
\noindent
\\
Rosenberg, J.L. \& Schneider, S.E. 2003, ApJ, 585, 256
\noindent
\\
Ryan-Weber, E.V., Webster, R.L., \& Stavely-Smith, L. 2003, MNRAS, 343, 1195
\noindent
\\
Sancisi, R., Fraternali, F., Oosterloo, T., \& van der Hulst, T. 2008, 
A\&ARv, 15, 189
\noindent
\\
Schaye, J., Carswell, R.F., \& Kim, T.-S. 2007, MNRAS, 379, 1169
\noindent
\\
Sembach, K.R., Howk, J.C., Savage, B.D., Shull, J.M. 2001,
AJ, 121, 992
\noindent
\\
Sembach, K.R., Savage, B.D., Lu, L., \& Murphy, E.M. 1999, ApJ, 515, 108
\noindent
\\
Sembach, K.R., Savage, B.D., Lu, L., \& 
Murphy, E.M. 1995, ApJ, 451, 616
\noindent
\\
Spergel, D.N., Bean, R., Dore\'e, O., et al.\, 2007, ApJS, 170, 377
\noindent
\\
Steidel, C.C., Kollmeier, J.A., Shapley, A.E., Churchill, C.W., Dickinson, M.
\& Pettini, M. 2002, ApJ, 570, 526
\noindent
\\
Stocke, J.T., Keeney, B.A., \& Danforth, C.W. 2010, PASA, 27, 256
\noindent
\\
Thilker, D.A., Braun, R., Westmeier, T. 2005, in: Extra-planar gas,
ASP Conference Series, 331, 113
\noindent
\\
Thilker, D.A., Braun, R., Walterbos, R.A.M., et al. 2004, ApJ, 601, L39
\noindent
\\
Thom, C., Peek, J.E.G., Putman, M.E., Heiles, C., Peek, K.M.G.,
\& Wilhelm, R. 2008, ApJ, 684, 364
\noindent
\\
Thom, C., Putman, M.E., Gibson, B.K., Christlieb, N., Flynn, C., 
Beers, T.C., Wilhelm, R., Lee, Y.S. 2006, ApJ, 638, L97
\noindent
\\
Turnshek, D.A., Rao, S., Nestor, D., Lane, W., Monier, E., 
Bergeron, J., \& Smette, A. 2001, ApJ, 553, 288
\noindent
\\
van\,de\,Voort, F., Schaye, J., Booth, C.M., \& Dalla\,Vecchia, C. 2011, 
MNRAS, 414, 2458
\noindent
\\
Vieser, W. \& Hensler, G. 2007, A\&A, 475, 251
\noindent
\\
Wakker, B.P., York, D.G., Wilhelm, R., Barentine, J.C., Richter, P.,
Beers, T.C., Ivezi\a'c, Z., \& Howk, J.C.\,2008, ApJ, 672, 298
\noindent
\\
Wakker, B.P., York, D.G., Howk, J.C., et al.\,2007, ApJ, 670, L113
\noindent
\\
Wakker, B.P. 2004, in ASSL, Vol.\,312,
High-Velocity Clouds, ed. van\,Woerden et al.
(Kluwer Academic Publishers), 251
\noindent
\\
Wakker, B.P. 2001, ApJS, 136, 463
\noindent
\\
Wakker, B.P., Howk, J.C., Savage, B.D. 1999, Nature, 402, 388
\noindent
\\
Wakker, B.P. \& van\,Woerden, H. 1998, ARA\&A, 35, 217
\noindent
\\
Weymann, R.J., Jannuzi, B.T., Lu, L., et al. 1998, ApJ, 506, 1
\noindent
\\
White, S.D.M. \& Frenk, C.S. 1991, ApJ, 379, 52
\noindent
\\
White, S.D.M. \& Rees, M.J. 1978, MNRAS, 183, 341
\noindent
\\
Winkel, B., Ben Bekhti, N., Darmst\"adter, V., Fl\"oer, L., Kerp, J., 
\& Richter, P. 2011, A\&A, 533, A105
\noindent
\\
Wolfe, A.M., Lanzetta, K.M., Foltz, C.B., \& Chaffee, F.H. 1995,
ApJ, 454, 698
\noindent
\\
Zwaan, M.A., van der Hulst, J.M., Briggs, F.H., Verheijen, M.A.W., \&
Ryan-Weber, E.V. 2005, MNRAS, 364, 1467

\end{small}

\end{document}